\documentclass[seceq,preprint]{ptptex}

\usepackage{graphicx}

\newcommand{\nn}{\nonumber\\}
\newcommand{\bea}{\begin{eqnarray}}
\newcommand{\ena}{\end{eqnarray}}
\newcommand{\vs}[1]{\vspace{#1 mm}}
\newcommand{\hs}[1]{\hspace{#1 mm}}
\renewcommand{\a}{\alpha}

\renewcommand{\c}{\gamma}
\newcommand{\s}{\sigma}
\newcommand{\pa}{\partial}
\newcommand{\p}[1]{(\ref{#1})}

\preprintnumber[3cm]{KU-TP 014
}

\title{Accelerating Cosmologies in the Einstein-Gauss-Bonnet Theory with Dilaton
}

\author{Kazuharu \textsc{Bamba},\footnote{e-mail address: bamba@phys.kindai.ac.jp}
Zong-Kuan \textsc{Guo}\footnote{e-mail address: guozk@phys.kindai.ac.jp}
and Nobuyoshi \textsc{Ohta}\footnote{e-mail address: ohtan@phys.kindai.ac.jp}
}

\inst{
Department of Physics, Kinki University, Higashi-Osaka 577-8502, Japan
}

\abst{
We study cosmological solutions in the low-energy effective heterotic string theory,
which is the Einstein gravity with Gauss-Bonnet term and the dilaton.
We show that the field equations are cast into an autonomous system for flat
internal and external spaces, and derive all the fixed points in the system.
We also examine the time evolution of the solutions and whether the solutions
can give (transient) accelerated expansion of our four-dimensional space
in the Einstein frame.
}

\begin{document}

\maketitle

\section{Introduction}

The recent cosmological observations have confirmed the existence
of the early inflationary epoch and the accelerated expansion of
the present universe~\cite{wmap}. An important problem is then
to derive such a model from fundamental theories of particle physics.
The most promising candidates for such theories are the ten-dimensional
superstrings or eleven-dimensional M-theory, which are hoped to give models of
accelerated expansion of the universe upon compactification to four dimensions.
There are many attempts to derive such models, but most of them assume some
additional matters or need special settings. From the viewpoint of the fundamental
theories, however, it is desirable if such models are obtained without
making special assumptions.

It has been shown that a model with certain period of accelerated expansion
can be obtained from the higher-dimensional vacuum Einstein equation
if one assumes a time-dependent hyperbolic internal space~\cite{tow03}
and that this class of models is obtained~\cite{oht03} from
what are known as S-branes~\cite{Wohlfarth:2003ni,Sbrane1} in the limit
of vanishing flux of three-form fields (see also Ref.~\citen{Sbrane3}). For other
attempts at inflation in the context of string theories, see, for
instance, Refs.~\citen{other,chen,Kachru}.
Unfortunately this class of models do not give sufficient inflation necessary
to resolve the cosmological problems.

On the other hand, it has been known that higher order corrections can give rise to
inflationary solutions.~\cite{starob} This is a very desirable setting since there
are terms of higher orders in the curvature to the lowest effective supergravity
action coming from superstrings or M-theory~\cite{Be,hetero0,hetero,Mth}.
The simplest such correction is the  Gauss-Bonnet (GB) term in the low-energy
effective heterotic string. (We ignore other gauge fields and forms for simplicity.)
It is thus important to examine what kind of time-dependent solutions are possible
in these theories.

There are many works discussing cosmology with the GB correction in four and
higher dimensions (see, for instance, \citen{Ish,GB1,GB2,GB3,Guo}).
For example, it was shown that there are two exponentially expanding solutions
in the higher-dimensional space, which may be called generalized de Sitter solutions
since the size of the internal space also depends on time.~\cite{Ish}
Note that this does not mean that the solutions gives accelerating expansion
in four dimensions.
Another interesting claim is that it is possible to obtain an inflationary
solutions if the coefficient of the Gauss-Bonnet term is negative,~\cite{GB2}
which is not the case in the effective theory of the heterotic string, and hence
may not be relevant in our consideration.
Moreover most of the work considers pure GB term without dilaton or assumes
constant dilaton, which is not the effective theory of the heterotic string, and
does not discuss cosmological solutions with dynamical dilaton in higher dimensions.
It is thus important to analyze the system including the dynamical dilatons.
Some attempt to obtain inflationary solutions in M theory with
higher order quantum corrections has also been made~\cite{MO}.

Recently a more interesting approach is considered for Einstein theory
with some additional scalars.~\cite{AH,TS}
In this dynamical system method, one considers the solution space restricted
by the constraint equation resulting from a component of the Einstein equation.
If the field equations are written as an autonomous system, we can
find fixed points in this space. Then all possible solutions are expressed
as trajectories between these fixed points in the solution space. This is
a very powerful method to examine possible solutions which is applicable
even if exact solutions are not available. In particular, it is possible
to find solutions with (transient) accelerating expansion which may be
relevant to cosmology. In fact, the existence of an eternally accelerating
solution, first found in Ref.~\citen{chen}, is established for hyperbolic internal
and external spaces without giving explicit solution.~\cite{AH}

In this paper, we consider cosmological solutions with a dilaton field
and the GB correction from heterotic string theory by extending the above
dynamical system method. We find that the field equations may be cast into
an autonomous system for flat internal and external spaces for both theories with
and without dynamical dilaton. We derive all the fixed points and analyze
their stability in the system.
We also examine the time evolution of the solutions and investigate whether
the solutions can give (transient) accelerated expansion of our
four-dimensional space in the Einstein frame.

This paper is organized as follows.
In \S~\ref{FE}, we first write down the action of the Einstein and GB theory,
and our metric for $D$-dimensional space. We then summarize the field equations.
In \S~\ref{nod}, we analyze the theory without the dilaton and find the solution space
and accelerating solutions. We show that the field equations become
an autonomous system, and find fixed points. We can see how the solutions
evolve in time by looking at the solution space and the flow. We find that
there is one fixed point corresponding to expanding solution with acceleration, but
it gives a singular super-inflation.
In \S~\ref{ddil}, we extend the analysis to the theory with the dilaton.
At first sight, the field equations cannot be reduced to an autonomous system,
but judicious choice of the time variable enables us to do it. We then
discuss their flow and properties of the fixed points of our system.
\S~\ref{conc} is devoted to conclusions.

\section{Field equations}
\label{FE}

We consider the low-energy effective action for the heterotic string:
\bea
S=\frac{1}{2\kappa_D^2}\int d^Dx \sqrt{-\tilde g}\, e^{-2\tilde
\phi} \left[\tilde R + 4(\partial_\mu \tilde \phi)^2 + \alpha_2
\tilde R^2_{\rm GB}\right],
\label{stringf}
\ena
where $\kappa_D^2$ is a $D$-dimensional gravitational constant,
$\tilde \phi$ is a dilaton field, $\alpha_2=\a'/8$ is a numerical
coefficient given in terms of the Regge slope parameter, and
$\tilde R^2_{\rm GB}=\tilde R_{\mu\nu\rho\sigma}\tilde R^{\mu\nu\rho\sigma}
- 4\tilde R_{\mu\nu}\tilde R^{\mu\nu} + \tilde R^2$ is the GB
correction. In the Einstein frame the dilaton $\tilde \phi$ is
minimally coupled to the metric and has a canonical kinetic term
\bea
S=\frac{1}{2\kappa_D^2}\int d^Dx \sqrt{-g} \left[R - \frac12 (\partial_\mu \phi)^2
+ \a_2 e^{-\c \phi} R^2_{\rm GB} \right],
\label{einsteinf}
\ena
where $g_{\mu\nu} = e^{-4 \tilde \phi/(D-2)}\tilde g_{\mu\nu}$,
$\phi = \sqrt{8/(D-2)}\, \tilde \phi$ and $\c=\sqrt{2/(D-2)}$.
Let us consider the metric in $D$-dimensional space
\bea
ds_D^2=-e^{2u_0(t)}dt^2 + e^{2u_1(t)}ds_p^2 + e^{2u_2(t)}ds_q^2 \,,
\ena
where $D=1+p+q$. The external $p$- and internal $q$-dimensional spaces
($ds_p^2$ and $ds_q^2$) are chosen to be maximally symmetric with the signature
of the curvature given by $\sigma_p$ and $\sigma_q$, respectively.
Though we are mainly concerned with flat internal and external spaces in
this paper, it may be useful to give field equations for more general case.

We find that the Riemann tensors are given by
\bea
R^t{}_{itj} &=& e^{-2u_0}X g_{ij} \,, \nn
R^t{}_{atb} &=& e^{-2u_0}Y g_{ab} \,, \nn
R^i{}_{jkl} &=& e^{-2u_0}A_p (g^i{}_k g_{jl} -g^i{}_l g_{jk}) \,, \nn
R^i{}_{ajb} &=& e^{-2u_0} \dot u_1 \dot u_2 g^i{}_j g_{ab} \,, \nn
R^a{}_{bcd} &=& e^{-2u_0}A_q (g^a{}_c g_{bd} -g^a{}_d g_{bc}) \,,
\ena
where $i,j$ and $a,b$ run over $p$- and $q$-dimensional spaces, respectively,
and
\bea
A_p&\equiv&\dot{u}_1^2+\sigma_p e^{2(u_0-u_1)}, \quad
A_q \equiv \dot{u}_2^2+\sigma_q e^{2(u_0-u_2)},\nn
X &\equiv& \ddot u_1 - \dot u_0 \dot u_1 +\dot u_1^2, \qquad
Y \equiv \ddot u_2 - \dot u_0 \dot u_2 +\dot u_2^2 \,.
\label{xy}
\ena
The GB term is given by
\bea
R^2_{\rm GB}
&=& e^{-4u_0} \Big\{ p_3 A_p^2 + 2p_1 q_1 A_p A_q + q_3 A_q^2
+ 4 \dot u_1 \dot u_2 (p_2 q A_p + p q_2 A_q) + 4 p_1 q_1 \dot u_1^2 \dot u_2^2 \nn
&& +\; 4pX \left[ (p-1)_2 A_p + q_1 A_q + 2(p-1)q \dot u_1 \dot u_2 \right]\nn
&& +\; 4qY \left[ p_1 A_p + (q-1)_2 A_q + 2p(q-1) \dot u_1 \dot u_2 \right]\Big\},
\label{gb1}
\ena
where we have defined
\bea
(p-m)_n&\equiv& (p-m)(p-m-1)(p-m-2)\cdots (p-n) \,,\nn
(q-m)_n&\equiv&(q-m)(q-m-1)(q-m-2)\cdots (q-n) \,,
\ena
Multiplying \p{gb1} by
$\sqrt{-g}e^{-\c \phi}=e^{u_0+pu_1+qu_2-\c \phi}$
and making partial integration,
one finds that the action reduces to the following (up to an overall factor):

\noindent
{\bf (1) Einstein-Hilbert action}
\bea
{\cal L}_1 =  e^{-u_0+pu_1+qu_2}\Bigl[p_1A_p + q_1A_q - 2(p_1\dot u_1{^2} +
pq\dot u_1\dot u_2 + q_1\dot u_2{^2} )+\frac12 \dot\phi^2\Bigr]\ .\label{eh2}
\ena
{\bf (2) GB action}
\bea
{\cal L}_2 &=& e^{-3u_0+pu_1+qu_2-\c\phi}\Bigl\{p_3A_p{^2} + 2p_1q_1A_qA_p+ q_3A_q{^2} \nn
&& -\; 4A_p (p_3\dot u_1{^2}+p_2q\dot u_1\dot u_2 + p_1q_1\dot u_2{^2})
 -4A_q(p_1q_1\dot u_1{^2}+pq_2\dot u_1\dot u_2 + q_3\dot u_2{^2})\nn
&& +\; \frac{4}{3}(2p_3\dot u_1{^4} + 2p_2q \dot u_1{^3}\dot u_2
+ 3p_1q_1\dot u_1{^2}\dot u_2{^2} +2pq_2\dot u_1\dot u_2{^3}
+ 2 q_3\dot u_2{^4}) \nn
&& +\; 4\c \dot \phi \Big[ (p_2 \dot u_1+p_1 q \dot u_2)A_p
+ (pq_1 \dot u_1+ q_2 \dot u_2)A_q - \frac23 \Bigl( p_2 \dot u_1^3 + q_2 \dot u_2^3
\Bigr)\Big] \Bigr\}\ .
\ena
If we set $\phi=0$, this agrees with the results in Ref.~\citen{MO}.

Now the field equations are
\bea
\label{fe1}
F &\equiv& F_1 + F_2 =0\,, \\
\label{fe2}
F^{(p)} &\equiv& f_1^{(p)} + f_2^{(p)} + X \left(g_1^{(p)} + g_2^{(p)}\right)
+ Y\left(h_1^{(p)} + h_2^{(p)}\right) - Z \;i^{(p)} =0\,, \\
\label{fe3}
F^{(q)} &\equiv& f_1^{(q)} + f_2^{(q)} + Y \left(g_1^{(q)} + g_2^{(q)}\right)
+ X\left(h_1^{(q)} + h_2^{(q)}\right) - Z \;i^{(q)} =0\,,\\
\label{fe4}
F_\phi &\equiv& Z + \a_2 \c e^{2u_0-\c \phi} R_{\rm GB}^2 =0\,,
\ena
where $R^2_{\rm GB}$ is given in Eq.~\p{gb1} and
\bea
Z &=& \ddot\phi +(-\dot u_0 + p\dot u_1 + q\dot u_2)\dot\phi\,, \nn
F_1&=& p_1 A_p+q_1 A_q+2pq\dot{u}_1\dot{u}_2 - \frac12 \dot\phi^2\,, \nn
f_1^{(p)}&=& (p-1)_2A_p+q_1 A_q+2(p-1)q\dot{u}_1\dot{u}_2 +\frac12 \dot\phi^2\,, \nn
f_1^{(q)}&=& p_1 A_p+(q-1)_2A_q+2p(q-1)\dot{u}_1\dot{u}_2
+\frac12 \dot\phi^2\,, \nn
g_1^{(p)}&=&2(p-1)\,, \nn
g_1^{(q)}&=&2(q-1)\,, \nn
h_1^{(p)}&=&2q\,, \nn
h_1^{(q)}&=&2p\,,
\label{eh1}
\ena
and
\bea
&&
F_2 = \a_2 e^{-2u_0-\c\phi} \Big\{ p_3 A_p^2+2p_1q_1 A_pA_q
+q_3 A_q^2 + 4(p_2 q A_p+p q_2 A_q + p_1q_1\dot{u}_1 \dot{u}_2)\dot{u}_1\dot{u}_2 \nn
&& \hs{10}
-\; 4 \c\dot\phi \big[ (p_2 \dot u_1 +p_1 q \dot u_2)A_p
+ (p q_1 \dot u_1 + q_2 \dot u_2)A_q + 2( p_1 q \dot u_1 + pq_1 \dot u_2) \dot u_1 \dot u_2 \big] \Big\}, \nn
&& f_2^{(p)}\; =\; \a_2 e^{-2u_0-\c\phi}\Big\{
(p-1)_4A_p^2+2(p-1)_2q_1 A_pA_q+q_3 A_q^2 \nn
&& \hs{20}
+\; 4\left[(p-1)_3qA_p+(p-1)q_2 A_q+(p-1)_2q_1\dot{u}_1\dot{u}_2
\right]\dot{u}_1\dot{u}_2 \nn
&& \hs{20}
+\; 4 \c  \dot \phi \big[ ((p-1)_2 A_p + q_1 A_q + 2(p-1)q \dot u_1 \dot u_2 )
(\dot u_1 + \c \dot\phi) \nn
&& \hs{20}
+\; 2 ( (p-1)_2 \dot u_1 A_p + q_1 \dot u_2 A_q +(p-1)q \dot u_1 \dot u_2
(\dot u_1 +\dot u_2) ) \big] \Big\}, \nn
&& f_2^{(q)}\; =\; \alpha_2 e^{-2u_0-\c\phi}\Big\{
p_3 A_p^2+ 2p_1 (q-1)_2 A_pA_q+(q-1)_4A_q^2 \nn
&& \hs{20}
+\; 4 \left[ p_2 (q-1)A_p+p(q-1)_3A_q+p_1(q-1)_2\dot{u}_1 \dot{u}_2
 \right]\dot{u}_1\dot{u}_2\nn
&& \hs{20}
+\; 4 \c  \dot \phi \big[ ( p_1 A_p + (q-1)_2 A_q + 2p(q-1) \dot u_1 \dot u_2 )
 (\dot u_2 + \c \dot\phi) \nn
&& \hs{20}
+\; 2 ( p_1 \dot u_1 A_p +(q-1)_2 \dot u_2 A_q + p(q-1) \dot u_1 \dot u_2
(\dot u_1 +\dot u_2) ) \big] \Big\}, \nn
&& g_2^{(p)}\; =\; 4(p-1)\alpha_2 e^{-2u_0-\c\phi}\Big[
(p-2)_3A_p+q_1A_q+2(p-2)q\dot{u}_1\dot{u}_2
-2 \c ((p-2)\dot u_1 +q \dot u_2) \dot \phi\Big],\nn
&& g_2^{(q)}\; =\; 4(q-1)\alpha_2 e^{-2u_0-\c\phi}\Big[
p_1A_p+(q-2)_3A_q+2p(q-2)\dot{u}_1\dot{u}_2
-2 \c (p\dot u_1 +(q-2) \dot u_2) \dot \phi\Big],\nn
&& h_2^{(p)}\; =\; 4q\alpha_2 e^{-2u_0-\c\phi}\Big[
(p-1)_2A_p+(q-1)_2A_q+2(p-1)(q-1)\dot{u}_1\dot{u}_2 \nn
&& \hs{30}
-\; 2 \c ((p-1)\dot u_1 +(q-1) \dot u_2) \dot \phi\Big], \nn
&& h_2^{(q)}\; =\; 4p\alpha_2 e^{-2u_0-\c\phi}\Big[
(p-1)_2A_p+(q-1)_2A_q+2(p-1)(q-1)\dot{u}_1\dot{u}_2 \nn
&& \hs{30}
-\; 2 \c ((p-1)\dot u_1 +(q-1) \dot u_2) \dot \phi\Big], \nn
&& i^{(p)}\; =\; \alpha_2 e^{-2u_0-\c\phi} 4 \c \Big[(p-1)_2 A_p
+ q_1 A_q + 2(p-1)q \dot u_1 \dot u_2\Big], \nn
&& i^{(q)}\; =\; \alpha_2 e^{-2u_0-\c\phi} 4 \c \Big[p_1 A_p + (q-1)_2 A_q
+ 2p(q-1) \dot u_1 \dot u_2\Big],
\label{gb2}
\ena

The basic Eqs.~\p{fe1} -- \p{fe4} are not all independent. They satisfy
\bea
\dot F +(p \dot u_1 + q \dot u_2 -2 \dot u_0) F
= p \dot u_1 F^{(p)} + q \dot u_2 F^{(q)} - \dot \phi F_\phi\,.
\ena
We are now going to examine cosmological solutions in this system.
In this paper, we only consider flat internal and external spaces, i.e.,
$\s_p=\s_q=0$. Henceforth, we set $p=3$ and $q=6$ though we write
formulae for more general cases as much as possible.

\section{Solutions in Einstein and Gauss-Bonnet theory}
\label{nod}

In this section, let us first consider the theory without dilaton
(or the case when dilaton is constant) as a consistency check.
Namely we set $\phi=0$ by hand to investigate the system without the dilaton field,
and study possible cosmological solutions in Einstein
and Gauss-Bonnet theory without dilaton.
This is the system examined in Ref.~\citen{Ish}, but the question
whether the accelerating expansion occurs or not in the four-dimensional spacetime
was not examined, and we clarify this point by making systematic analysis of
the solutions by the dynamical system method.
We can set $u_0=0$ by using time-reparametrization invariance.
It is also possible to put $\a_2=1$ by choosing a suitable unit of time.~\cite{MO}

\begin{figure}[tbhp]
\setlength{\unitlength}{.7mm}
\begin{center}
\begin{picture}(30,75)(65,0)
\includegraphics[width=11cm]{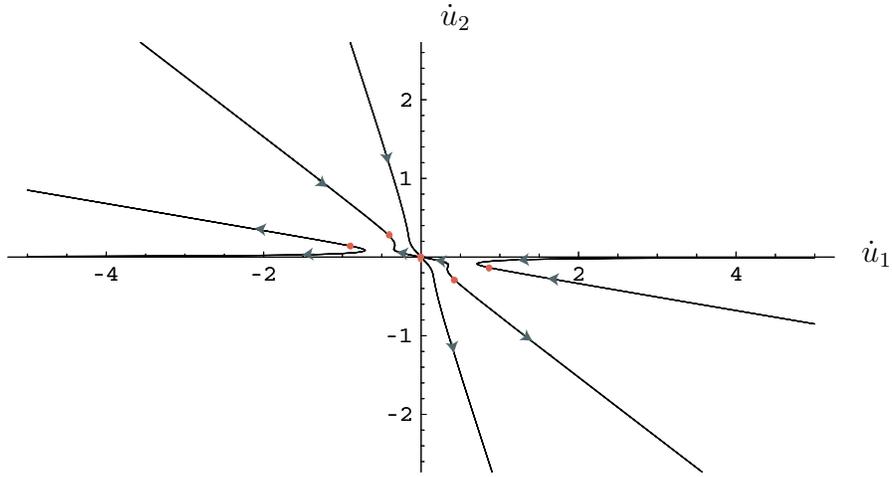}
\put(5,40){$\dot u_1$}
\put(-75,85){$\dot u_2$}
\end{picture}
\caption{Solution space and flow. The dots indicate fixed points.}
\label{solnd}
\end{center}
\end{figure}

Equation~\p{fe1} is a constraint equation, and any cosmological solutions
must satisfy this. In this sense, this gives the space in which all the possible
solutions live, which we call solution space. This is depicted in Fig.~\ref{solnd}
in $(\dot u_1, \dot u_2)$-plane.

Solving Eqs.~\p{fe2} and \p{fe3} for $\ddot u_1$ and $\ddot u_2$,
we find that the field equations~\p{fe2} and \p{fe3}
become an autonomous system for $\dot u_1$ and $\dot u_2$.
We then find the five fixed points of these variables for $p=3,q=6$ in the unit
of $\a_2=1$ are given by
\bea
(\dot u_1, \dot u_2)~ =~ (0,0), ~~
(\pm 0.88603, \mp 0.13845), ~~
(\pm 0.48296, \mp 0.34141),
\label{fix}
\ena
which are also shown in Fig.~\ref{solnd}.
We can also derive the flow of the solutions along the time lapse between
the fixed points as shown in the figure. Due to the time-reversal symmetry
of the system, the figure is symmetric under $\pi$ rotation (with the reversed
time flow). All this agrees with the results in Ref.~\citen{Ish}.

Our cosmological model is higher-dimensional, and there are two kinds of
frames that we can take to discuss cosmologies, the original frame and
the Einstein frame in four dimensions.
Note that this Einstein frame is different from the one defined in Eq.~\p{einsteinf}
with respect to the dilaton. Instead it is a new frame which is defined to
eliminate the scalar fields which appear from the internal space by Kaluza-Klein
compactification to the external space. This is the frame in which the Newton
constant is really constant.
We must determine which frame is important for a successful
inflationary scenario. Since flatness and horizon problems should be
explained in our four-dimensional spacetime, that is, in the Einstein frame,
we should require a successful inflation in the Einstein frame.

\begin{figure}[htb]
\setlength{\unitlength}{.7mm}
\begin{center}
\begin{picture}(80,100)(45,0)
\includegraphics[width=15cm]{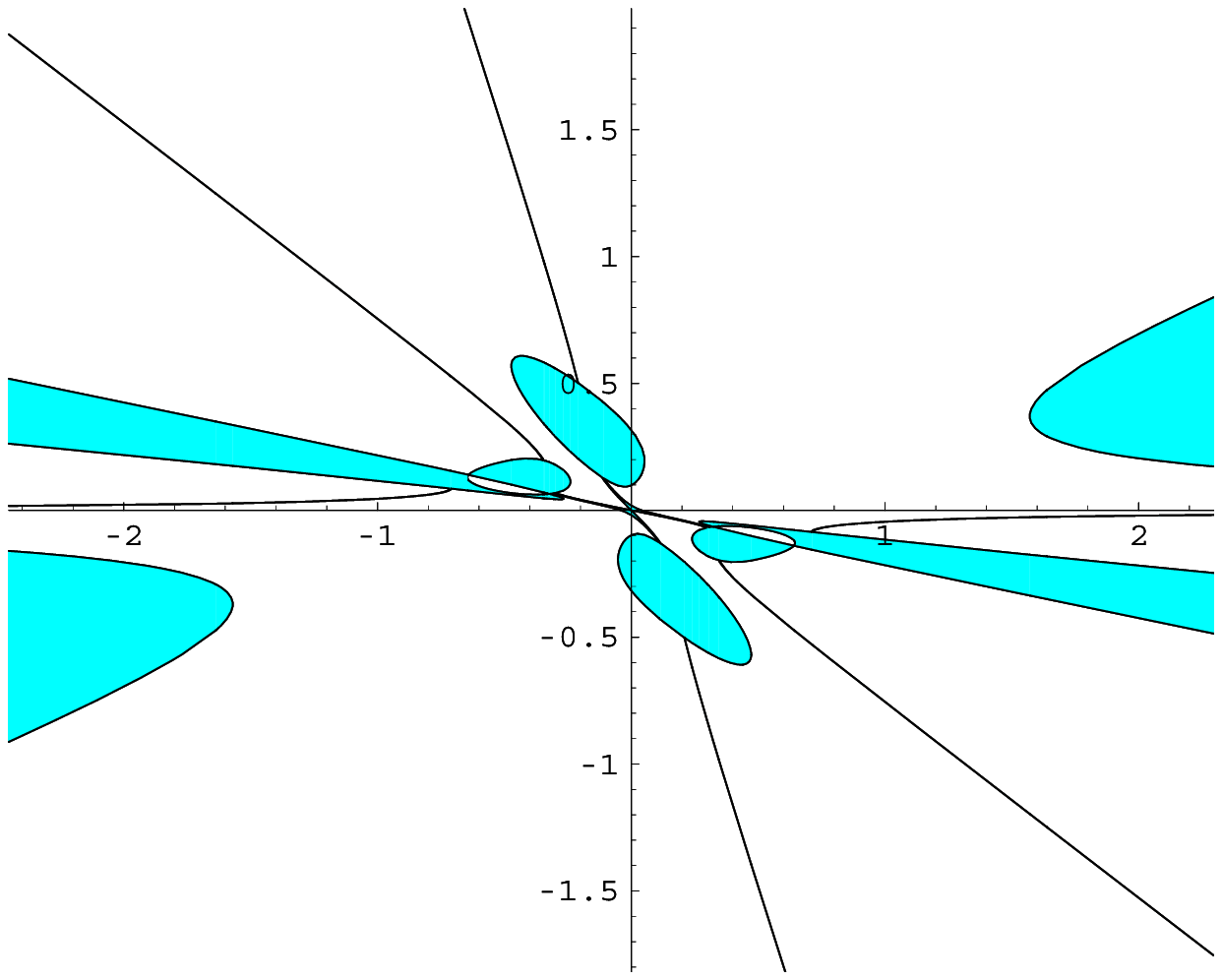}
\put(-55,45){\footnotesize $\dot u_1$}
\put(-120,105){\footnotesize $\dot u_2$}
\end{picture}
\caption{accelerating region in the solution space.}
\label{solaccnd}
\end{center}
\begin{center}
\begin{picture}(50,100)(30,0)
\includegraphics[width=9cm]{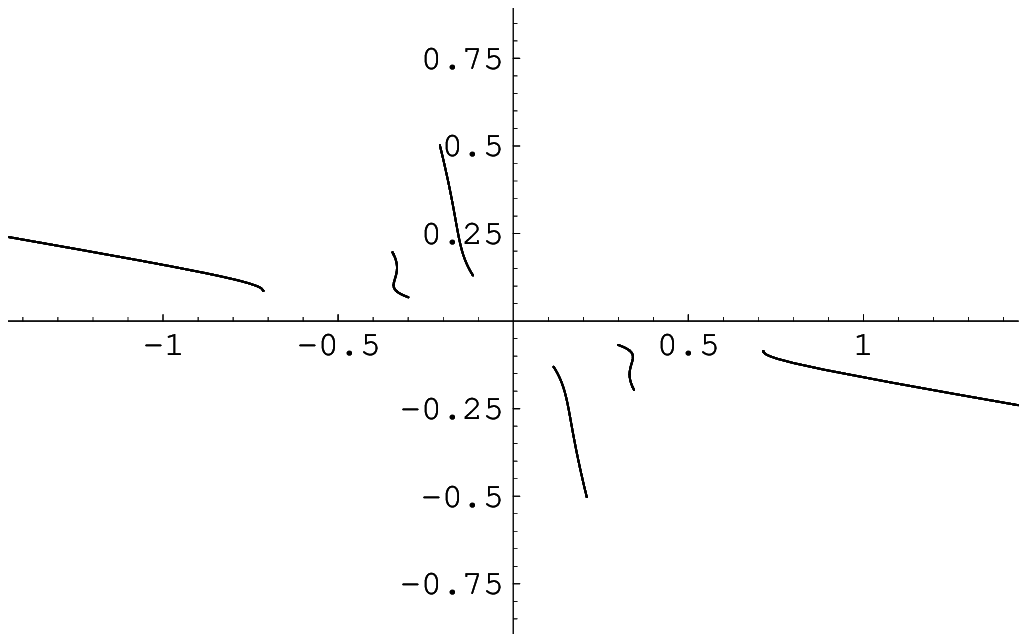}
\put(5,40){\footnotesize $\dot u_1$}
\put(-65,85){\footnotesize $\dot u_2$}
\end{picture}
\vs{5}
\caption{accelerating solutions.}
\label{solaccndf}
\end{center}
\end{figure}

Now let us examine if there is any region where the accelerating expansion is
realised in the four-dimensional Einstein frame.
The Einstein frame is obtained by
\bea
ds^2_D = e^{-\frac{2q}{p-1} u_2} ds_E^2 +e^{2u_2}ds_q^2.
\ena
So
\bea
ds_E^2 &=& e^{\frac{2q}{p-1} u_2} (-dt^2 + e^{2u_1}ds_p^2) \nn
&=& -d\tau^2 + a^2(\tau) ds_p^2,
\ena
where we have defined the cosmic time $\tau$ and scale factor by
\bea
\frac{d\tau}{dt}= e^{\frac{q}{p-1} u_2}, \quad
a(\tau) = e^{u_1 + \frac{q}{p-1} u_2}.
\ena
For $p=3,q=6$, the condition for expansion is
\bea
\frac{da}{d\tau}=\frac{dt}{d\tau} \frac{da}{dt} = (\dot u_1+3\dot u_2) e^{u_1}>0,
\label{expansion}
\ena
and the condition for accelerated expansion is
\bea
\frac{d^2a}{d\tau^2}&=&\frac{dt}{d\tau} \frac{d}{dt}\Big((\dot u_1+3\dot u_2)e^{u_1}
\Big) \nn
&=& \{ \ddot u_1+3\ddot u_2 + (\dot u_1+3\dot u_2)\dot u_1 \} e^{u_1-3u_2}>0.
\label{accelerat}
\ena
Substituting $\ddot u_1$ and $\ddot u_2$ into \p{accelerat},
we find that the accelerating regions are those shown in Fig.~\ref{solaccnd}, and
the solution space in these regions are depicted in Fig.~\ref{solaccndf}.

We have also examined the stability of the fixed points.
We find that the fixed points, $(0,0)$, corresponding to the flat
Minkowski space, $(-0.88603, 0.13845)$ and $(0.48296, -0.34141)$,
corresponding to a contracting universe, are unstable. The fixed
points $(-0.48296, 0.34141)$ and $(0.88603,- 0.13845)$ are stable.
Among these, only the last one gives accelerating expansion.
The behavior of the scale factor is like $a \propto
|\tau|^{-1.13}$ for negative $\tau$. This is what is called
super-inflation and exhibits singularity near $\tau \sim 0$.
There is a solution flowing into this fixed point which exhibits accelerating
expansion for its whole evolution.
Such a super-inflation may also arise in phantom cosmological models~\cite{pia03}.
This is also called a Big Rip and should be avoided~\cite{MO}.
However, there are several (transient) accelerating cosmological solutions.
For example, there is a solution coming out of the flat Minkowski space
flowing into the direction of positive $\dot u_1$ and negative $\dot u_2$ with
transient acceleration. Whether this solution gives viable cosmological solution
or not remains to be examined.

\section{Solutions with a dynamical dilaton}
\label{ddil}

In this section, we extend our analysis to the more interesting theory with
a dynamical dilaton with $p=3$ and $q=6$, which appears as a low-energy effective
theory of the heterotic string.
We note again that $u_0=0$ can be chosen by time reparametrization,
and the choice of a suitable unit of time can be used to set $\a_2=1$~\cite{MO}.

In this system, there are exponential factors of the dilaton in the
field equations~\p{fe1} -- \p{fe4}, and this appears to prevent us from writing
them as an autonomous system. However, if we introduce new time variable $T$ by
\bea
\pa_t = e^{\phi/4} \pa_T, \quad
i.e. \quad
\frac{dT}{dt}=e^{\phi/4},
\ena
then it is possible to rewrite them as an autonomous system.
In what follows, derivatives with respect to $T$ will be denoted by the prime $'$.
Then the field equations~\p{fe1} -- \p{fe4} remain the same if we make
the following replacement in Eqs.~\p{xy}, \p{eh1} and \p{gb2}:
\bea
\ddot u_1 \to u_1''+\frac14 u_1' \phi', \quad
\ddot u_2 \to u_2''+\frac14 u_2' \phi', \quad
\ddot \phi \to \phi''+\frac14 (\phi')^2,
\label{rep}
\ena
and remove the exponential factors in Eqs.~\p{fe4} and \p{gb2}.

\begin{figure}[htb]
\begin{minipage}{.5\linewidth}
\begin{center}
\includegraphics[width=7cm]{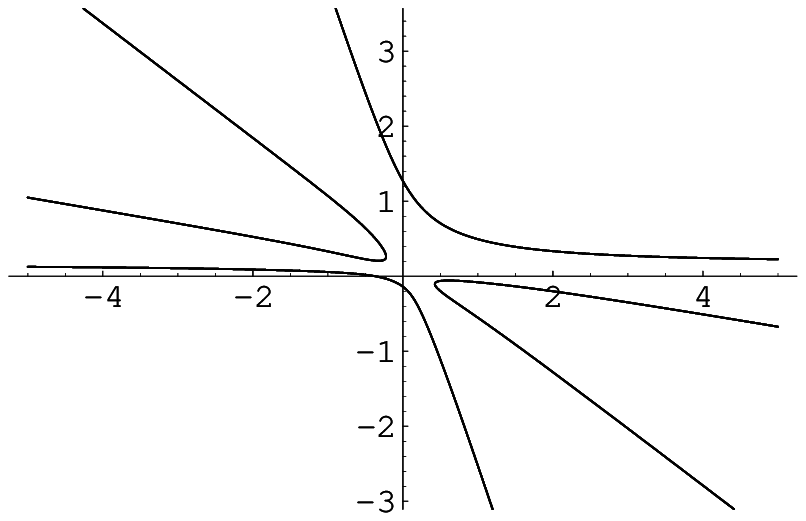}
\end{center}
\caption{Slice at $\phi'=2$ in the solution space.}
\label{sold1}
\begin{center}
\includegraphics[width=7cm]{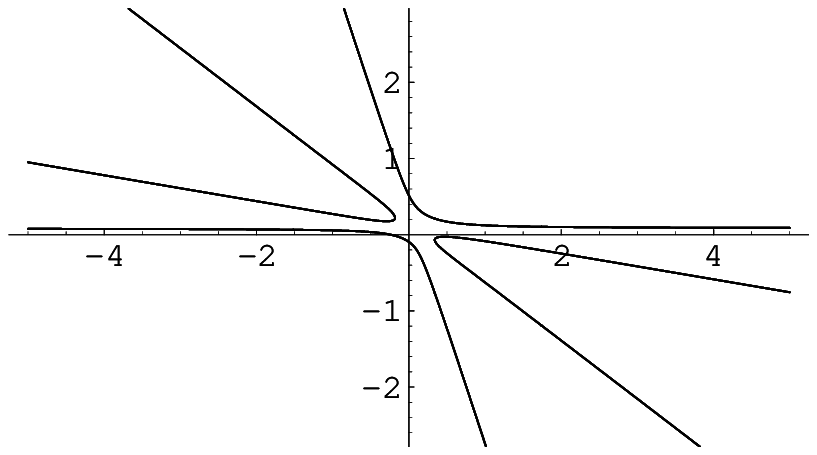}
\end{center}
\caption{Slice at $\phi'=1$ in the solution space.}
\label{sold2}
\begin{center}
\includegraphics[width=7cm]{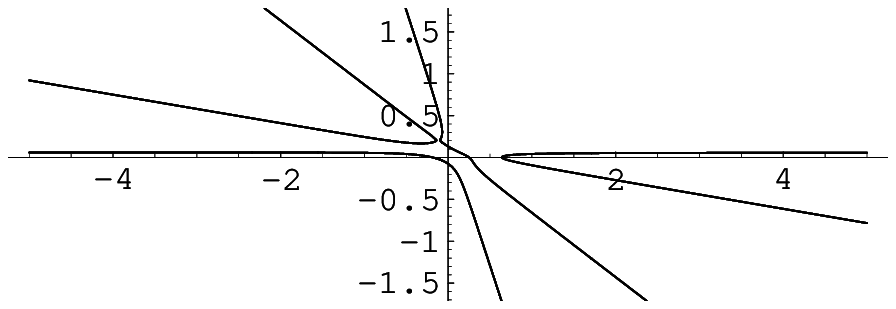}
\end{center}
\caption{Slice at $\phi'=0.7$ in the solution space.}
\label{sold3}
\end{minipage}
\begin{minipage}{.5\linewidth}
\begin{center}
\includegraphics[width=7cm]{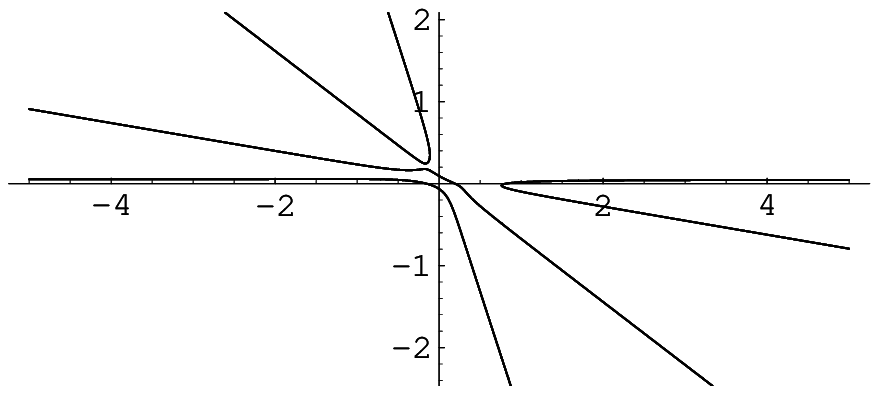}
\end{center}
\caption{Slice at $\phi'=0.5859$ in the solution space.}
\label{sold4}
\vs{2}
\begin{center}
\includegraphics[width=7cm]{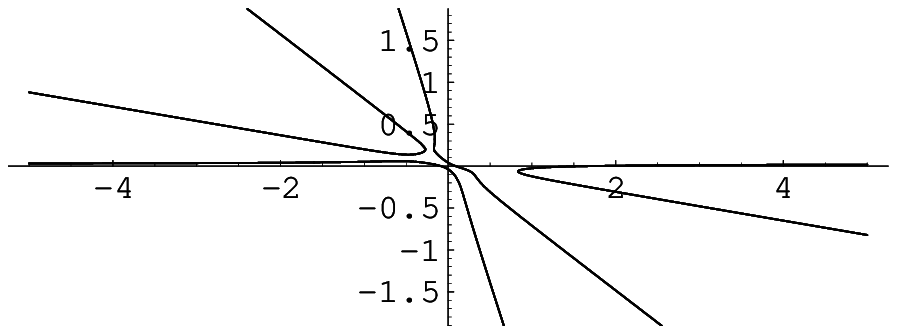}
\end{center}
\caption{Slice at $\phi'=0.3$ in the solution space.}
\label{sold5}
\begin{center}
\includegraphics[width=6cm]{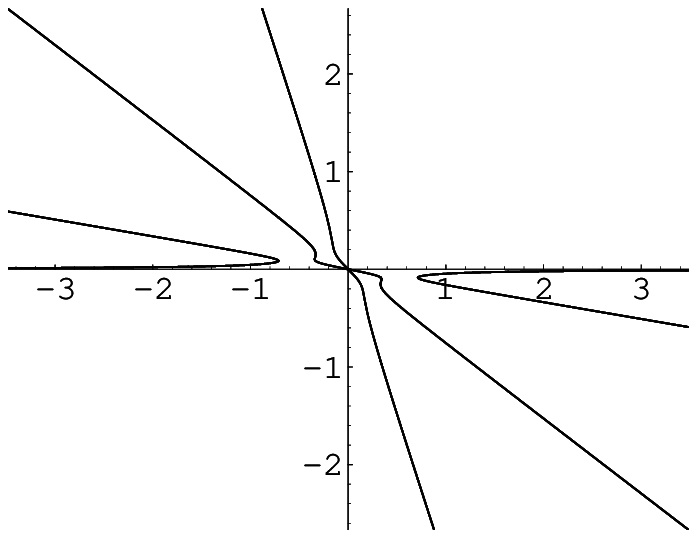}
\end{center}
\caption{Slice at $\phi'=0$ in the solution space.}
\label{sold6}
\end{minipage}
\end{figure}

This is again an autonomous system for $x\equiv u_1'$, $y\equiv u_2'$ and
$z\equiv \phi'$. Among these, the constraint~\p{fe1} gives the solution space.
In this case, because we have 3 variables $x$, $y$ and $z$,
the space consists of 2-dimensional surfaces embedded in 3 dimensions.
The surfaces have the shape of hyperbolic surfaces.
Since it does not seem to be so instructive to draw the surface in 3 dimensions,
we show the shapes of slices of the solution space at $\phi'=2, 1,0.7, 0.585906,0.3,0$
in Figs.~\ref{sold1} -- \ref{sold6}, respectively.
We see that reconnections of the surfaces occur as $\phi'$ varies.
Note also that the region for $\phi'<0$ has just the $\pi$-rotated shape
due to time reversal symmetry.

We find that there are seven fixed points in this system
\bea
(x, y, z) &=& \mbox{M}  (0,0,0), ~~
\mbox{P}_1 (\mp 0.292373, \pm 0.36066, \pm 0.954846), \nn
&& \hs{-10}
\mbox{P}_2 (\pm 0.91822, \mp 0.080285, \pm 0.585906), ~~
\mbox{P}_3 (\pm 0.161307, \pm 0.161307, \mp 9.30437),~~
\ena
where the labels are indicated for upper signs and those lower signs are
denoted with tildes.
Their properties are summarized in Table~\ref{crit}.
\begin{table}
\begin{center}
\begin{tabular}{c|c|c|c|c|c} \hline\hline
Label & $(x,y,z)$ &  Eigenvalues $(\lambda_1, \lambda_2)$ &
Stability & $\frac{da}{d\tau}$ & $\frac{d^2a}{d\tau^2}$ \\
\hline \hline
M  & $(0,0,0)$ & $(0, 0)$ & unstable & -- & -- \\
\hline
P$_1$ & $(0.292373,-0.36066,-0.954846)$ & $(1.52555,1.52555)$ & unstable & $<0$ & $<0$ \\
\hline
$\tilde{\rm P}_1$ & $(-0.292373,0.36066,0.954846)$ & $(-1.52555,-1.52555)$ & stable
 & $>0$ & $<0$ \\
\hline P$_2$ & $(0.91822,-0.080285,0.585906)$ &
$(-2.41943,-2.41943)$ & stable
 & $>0$ & $>0$ \\
\hline $\tilde{\rm P}_2$ & $(-0.91822,0.080285,-0.585906)$ &
$(2.41943,2.41943)$ &
 unstable & $<0$ & $>0$ \\
\hline
P$_3$ & $(0.161307,0.161307,-9.30437)$ & $(0.874329,0.874324)$ & unstable & $>0$ & $<0$ \\
\hline
$\tilde{\rm P}_3$ & $(-0.161307,-0.161307,9.30437)$ & $(-0.87433,-0.874324)$ &
 stable & $<0$ & $<0$ \\
\hline\hline
\end{tabular}
\end{center}
\caption[crit]{Fixed points of the autonomous system and their properties.}
\label{crit}
\end{table}

The expansion criterion~\p{expansion} with replacement~\p{rep}
tells us that the solutions of $\tilde{\rm P}_1$, P$_2$ and P$_3$ give
the expanding solutions in the Einstein frame. Among these, only P$_2$ gives
accelerating expansion. In this accelerated solution, we have
\bea
T = \frac{4}{\phi'} e^{\frac{\phi'}{4} t}, \quad
\frac{d\tau}{dT}=e^{3u_2-\phi/4}= e^{-0.387 T}.
\ena
We then find that this solution with
\bea
a(\tau) = e^{u_1+3u_2}=e^{0.677 T} \sim |\tau|^{-1.75},
\ena
gives again a super-inflation and
$\tau$ changes from $-\infty$ to $0$ as $T$ changes from $-\infty$ to $\infty$.

In order to study the stability of the fixed points, we substitute linear
perturbations $x \to x + \delta x$, $y \to y + \delta y$ and $z \to z + \delta z$
about the fixed points into the field equations~\p{fe1} -- \p{fe4}.
To the first order in the perturbations,
we obtain two independent equations of motion which can be written as
\bea
\left(
\begin{array}{c}
\delta x' \\
\delta y'
\end{array}
\right) = {\cal M} \left(
\begin{array}{c}
\delta x \\
\delta y
\end{array}
\right),
\ena
where $\cal M$ is a $2 \times 2$ matrix. Stability requires that both
the eigenvalues of the matrix $\cal M$, $\lambda_1$ and $\lambda_2$
be negative. Our analysis shows that $M$, P$_1$, $\tilde{\rm P}_2$ and
P$_3$ are unstable while $\tilde{\rm P}_1$, P$_2$ and $\tilde{\rm P}_3$ are stable.

\begin{figure}[htb]
\begin{center}
\includegraphics[width=8cm]{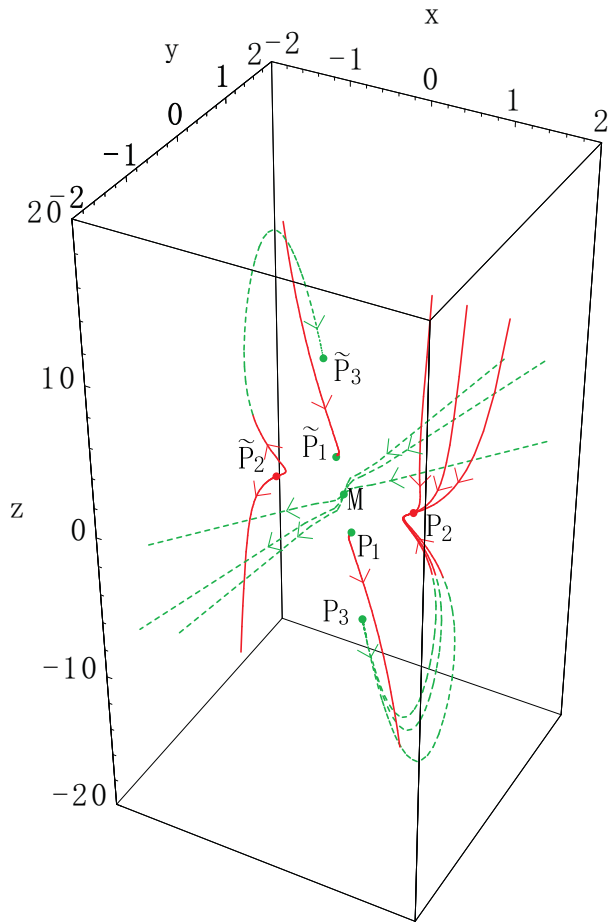}
\end{center}
\caption{Solution space and flow in the case with a dynamical dilaton.
The solid (red) lines correspond to $d^2a/d\tau^2>0$ and the dashed (green)
lines correspond to $d^2a/d\tau^2<0$.}
\label{fig10}
\end{figure}

The flow diagram for solutions around the fixed points is drawn in Fig.~\ref{fig10}.
We can use it to examine what kind of solutions are possible.
For example, there are solutions starting from a decelerated expanding
region which approach the accelerated expanding solution (P$_2$),
solutions starting from a decelerated contracting region which approach
the accelerated contracting solution ($\tilde{\rm P}_3$),
and solutions starting from a accelerated expanding region which approach
the decelerated expanding solution ($\tilde{\rm P}_1$).
We see from this figure that there are several accelerating cosmological solutions
in this theory including those flowing into non-accelerating fixed point
$\tilde{\rm  P}_1$ and those flowing into P$_2$ with Big Rip singularity.
It is possible that stringy effects resolve this kind of singularity and
these solutions may give viable cosmologies.

It is interesting to investigate whether these solutions give viable cosmological
solution or not.
A step towards this is to examine if we can get enough e-folding for solving
cosmological problems. A preliminary investigation of the solution flowing into
the fixed point P$_2$ indicates that it is hard to get enough
e-folding number before arriving at the fixed point but we can easily get
sufficient e-folding number if the solution arrives at the fixed point.

When we consider the accelerating expansion of the present
universe, the fine-tuning problem is always a nagging problem.
To partially answer this question, we have examined solutions by changing
initial conditions near the fixed point P$_2$, and find that there are several
solutions flowing into P$_2$, as shown in Fig.~\ref{fig10}.
This means that there are certain range of initial
conditions which lead to the accelerating expansion. In this sense, these solutions
have the possibility of explaining naturalness of the accelerating expansion.
To examine how large area of these initial conditions can give such a behavior and
whether the present model can give realistic one need further study.

\section{Conclusions}
\label{conc}

In this paper we have investigated cosmological solutions in the Einstein theory
with GB correction with and without a dilaton in higher dimensions.
We are interested in this theory because this is the low-energy effective
theory of the heterotic string, and examined what solutions are possible
by the dynamical system method.
For flat internal and external spaces, we have shown that the field equations
can be written as an autonomous system for both the theories with and without
dynamical dilaton. We obtained the fixed points and analyzed their stability.
We have found that both in the GB correction with and without dilaton,
there are solutions with accelerating expansion. Some of them are
super-inflation with future singularity.

The analysis in Ref.~\citen{AH} indicates that even if there is no interesting
cosmological solution in the Einstein theory for flat internal and external spaces,
there may exist an interesting solution with eternally accelerating expansion
for curved spaces. The existence of such a solution was originally suggested
in Ref.~\citen{chen} by a perturbation around non-inflationary solution, and it
was shown that the solution is eternally expanding with acceleration after some time.
However, due to the limitation of the perturbation, the detailed properties of
the solution (like its eternal accelerating property for whole time) was
not clear. The powerful method of dynamical system allowed to show that the
solution is eternally expanding with acceleration for the whole time.~\cite{AH}
It is thus possible that similarly interesting solutions may exist in our
Einstein-Gauss-Bonnet gravity coupled to dilaton.
It would be very interesting to extend our analysis to curved external
and internal spaces, and check if there may be additional interesting solutions.

\section*{Acknowledgments}

We would like to thank K. Maeda and S. Tsujikawa for useful cooespondence.
This work was supported in part by the Grant-in-Aid for
Scientific Research Fund of the JSPS Nos. 16540250 and 06042.
K.B. was also supported in part by the
open research center project at Kinki University.



\begin{thebibliography}{99}
\bibitem{wmap}
D.~N.~Spergel {\it et al.}  [WMAP Collaboration],
  Astrophys.\ J.\ Suppl.\  {\bf 148} (2003) 175
  [arXiv:astro-ph/0302209];
  Astrophys.\ J.\ Suppl.\  {\bf 170} (2007) 377
  [arXiv:astro-ph/0603449];\\
H.~V.~Peiris {\it et al.}  [WMAP Collaboration],
  Astrophys.\ J.\ Suppl.\  {\bf 148} (2003) 213
  [arXiv:astro-ph/0302225].

\bibitem{tow03}
P.~K.~Townsend and M.~N.~R.~Wohlfarth,
  Phys.\ Rev.\ Lett.\  {\bf 91} (2003) 061302
  [arXiv:hep-th/0303097].

\bibitem{oht03}
N.~Ohta,
  Phys.\ Rev.\ Lett.\  {\bf 91} (2003) 061303
  [arXiv:hep-th/0303238];
  Prog.\ Theor.\ Phys.\  {\bf 110} (2003) 269
  [arXiv:hep-th/0304172].

\bibitem{Wohlfarth:2003ni}
M.~N.~R.~Wohlfarth,
  Phys.\ Lett.\ B {\bf 563} (2003) 1
  [arXiv:hep-th/0304089].

\bibitem{Sbrane1}
C.~M.~Chen, D.~V.~Gal'tsov and M.~Gutperle,
  Phys.\ Rev.\ D {\bf 66} (2002) 024043
  [arXiv:hep-th/0204071];\\
N.~Ohta,
  Phys.\ Lett.\ B {\bf 558} (2003) 213
  [arXiv:hep-th/0301095].


\bibitem{Sbrane3}
L.~Cornalba and M.~S.~Costa,
  Phys.\ Rev.\ D {\bf 66} (2002) 066001
  [arXiv:hep-th/0203031];\\
C.~P.~Burgess, F.~Quevedo, S.~J.~Rey, G.~Tasinato and I.~Zavala,
  JHEP {\bf 0210} (2002) 028
  [arXiv:hep-th/0207104];\\
S.~Roy,
  Phys.\ Lett.\ B {\bf 567} (2003) 322
  [arXiv:hep-th/0304084];\\
A.~Buchel and J.~Walcher,
  JHEP {\bf 0305} (2003) 069
  [arXiv:hep-th/0305055];\\
C.~Armendariz-Picon and V.~Duvvuri,
  Class.\ Quant.\ Grav.\  {\bf 21} (2004) 2011
  [arXiv:hep-th/0305237];\\
C.~P.~Burgess, P.~Martineau, F.~Quevedo, G.~Tasinato and I.~Zavala C.,
  JHEP {\bf 0303} (2003) 050
  [arXiv:hep-th/0301122];\\
I.~P.~Neupane and D.~L.~Wiltshire,
  Phys.\ Lett.\ B {\bf 619} (2005) 201
  [arXiv:hep-th/0502003];
  Phys.\ Rev.\ D {\bf 72} (2005) 083509
  [arXiv:hep-th/0504135].
\bibitem{other}
L.~Cornalba and M.~S.~Costa,
  Fortsch.\ Phys.\  {\bf 52} (2004) 145
  [arXiv:hep-th/0310099];\\
V.~Balasubramanian,
  Class.\ Quant.\ Grav.\  {\bf 21} (2004) S1337
  [arXiv:hep-th/0404075];\\
N.~Ohta,
  Int.\ J.\ Mod.\ Phys.\ A {\bf 20} (2005) 1
  [arXiv:hep-th/0411230].

\bibitem{chen}
C.~M.~Chen, P.~M.~Ho, I.~P.~Neupane, N.~Ohta and J.~E.~Wang,
JHEP {\bf 0310} (2003) 058
[arXiv:hep-th/0306291];
JHEP {\bf 0611} (2006) 044
[arXiv:hep-th/0609043].

\bibitem{Kachru}
S.~Kachru, R.~Kallosh, A.~Linde, J.~M.~Maldacena, L.~McAllister and S.~P.~Trivedi,
  JCAP {\bf 0310} (2003) 013
  [arXiv:hep-th/0308055].

\bibitem{starob}
A.~A.~Starobinsky,
  Phys.\ Lett.\  B {\bf 91} (1980) 99.

\bibitem{Be}
D.~J.~Gross and J.~H.~Sloan,
  Nucl.\ Phys.\ B {\bf 291} (1987) 41.
\bibitem{hetero0}
M.~de Roo, H.~Suelmann and A.~Wiedemann,
  Nucl.\ Phys.\ B {\bf 405} (1993) 326
  [arXiv:hep-th/9210099].
\bibitem{hetero}
A.~A.~Tseytlin,
  Nucl.\ Phys.\ B {\bf 467} (1996) 383
  [arXiv:hep-th/9512081].
\bibitem{Mth}
K.~Peeters, P.~Vanhove and A.~Westerberg,
  Class.\ Quant.\ Grav.\  {\bf 18} (2001) 843
  [arXiv:hep-th/0010167].

\bibitem{Ish}
H.~Ishihara,
Phys.\ Lett.\  B {\bf 179} (1986) 217.

\bibitem{GB1}
K.~Maeda,
  Phys.\ Lett.\  B {\bf 166} (1986) 59;\\
B.~C.~Paul and S.~Mukherjee,
  Phys.\ Rev.\  D {\bf 42} (1990) 2595;\\
M.~Gasperini and M.~Giovannini,
  Phys.\ Lett.\  B {\bf 287} (1992) 56.

\bibitem{GB2}
M.~H.~Dehghani,
  Phys.\ Rev.\  D {\bf 70} (2004) 064009.

\bibitem{GB3}
S.~Nojiri, S.~D.~Odintsov and M.~Sasaki,
  Phys.\ Rev.\  D {\bf 71} (2005) 123509
  [arXiv:hep-th/0504052];\\
G.~Calcagni, S.~Tsujikawa and M.~Sami,
  Class.\ Quant.\ Grav.\ {\bf 22} (2005) 3977
  [arXiv:hep-th/0505193];\\
S.~Nojiri, S.~D.~Odintsov and M.~Sami,
  Phys.\ Rev.\  D {\bf 74} (2006) 046004
  [arXiv:hep-th/0605039];\\
S.~Tsujikawa,
  Annalen Phys.\ {\bf 15} (2006) 302
  [arXiv:hep-th/0606040];\\
T.~Koivisto and D.~F.~Mota,
  Phys.\ Lett.\  B {\bf 644} (2007) 104
  [arXiv:astro-ph/0606078];
  Phys.\ Rev.\  D {\bf 75} (2007) 023518
 [arXiv:hep-th/0609155];\\
K.~Andrew, B.~Bolen and C.~A.~Middleton,
  arXiv:hep-th/0608127;\\
S.~Tsujikawa and M.~Sami,
  JCAP {\bf 0701} (2007) 006
  [arXiv:hep-th/0608178]; \\
S.~Nojiri and S.~D.~Odintsov,
  arXiv:hep-th/0611071; \\
G.~Cognola {\it et al.},
  Phys.\ Rev.\  D {\bf 75} (2007) 086002
  [arXiv:hep-th/0611198];\\
E.~Elizalde {\it et al.},
  Phys.\ Lett.\  B {\bf 644} (2007) 1
  [arXiv:hep-th/0611213];\\
B.~M.~Leith and I.~P.~Neupane,
  JCAP {\bf 0705} (2007) 019
  [arXiv:hep-th/0702002];\\
L.~Amendola, C.~Charmousis and S.~C.~Davis,
  arXiv:0704.0175 [astro-ph];\\
A.~Sheykhi, B.~Wang and N.~Riazi,
  Phys.\ Rev.\ D {\bf 75} (2007) 123513
  [arXiv:0704.0666];\\
S.~Nojiri, S.~D.~Odintsov and P.~V.~Tretyakov,
  arXiv:0704.2520; \\
E.~Elizalde {\it et al.},
  arXiv:0705.1211; \\
F.~Canfora, A.~Giacomini and S.~Willison,
  arXiv:0706.2891.

\bibitem{Guo}
Z.~K.~Guo, N.~Ohta and S.~Tsujikawa,
  Phys.\ Rev.\  D {\bf 75} (2007) 023520
  [arXiv:hep-th/0610336].

\bibitem{MO}
K.~Maeda and N.~Ohta,
Phys.\ Lett.\  B {\bf 597} (2004) 400
[arXiv:hep-th/0405205];
Phys.\ Rev.\  D {\bf 71} (2005) 063520
[arXiv:hep-th/0411093];\\
K.~Akune, K.~Maeda and N.~Ohta,
Phys.\ Rev.\  D {\bf 73} (2006) 103506
[arXiv:hep-th/0602242].

\bibitem{AH}
L.~Andersson and J.~M.~Heinzle,
  arXiv:hep-th/0602102.
\bibitem{TS}
J.~Sonner and P.~K.~Townsend,
  Phys.\ Rev.\  D {\bf 74} (2006) 103508
  [arXiv:hep-th/0608068].

\bibitem{pia03}
Y.~S.~Piao and E.~Zhou,
  Phys.\ Rev.\  D {\bf 68} (2003) 083515
  [arXiv:hep-th/0308080];\\
Y.~S.~Piao and Y.~Z.~Zhang,
  Phys.\ Rev.\  D {\bf 70} (2004) 063513
  [arXiv:astro-ph/0401231];\\
Z.~K.~Guo, Y.~S.~Piao and Y.~Z.~Zhang,
  Phys.\ Lett.\  B {\bf 594} (2004) 247
  [arXiv:astro-ph/0404225].

\end{thebibliography}
\end{document}